# Studies of Copper Nanoparticles Effects on Micro-organisms

T. Theivasanthi [*] and M. Alagar

*Centre for Research and P.G.Department of Physics, Ayya Nadar Janaki Ammal College, Sivakasi-626124, Tamilnadu, India.*

**\*Corresponding author.**     *E-mail*: theivasanthi@pacrpoly.org

___________________________________________________________________________

**Abstract**

*We discuss about the antibacterial activities of copper nanoparticles on both Gram negative and Gram positive bacteria in this investigation. First time, we increase its antibacterial activities by using electrical power while on electrolysis synthesis and it is confirmed from its more antibacterial activities (For Escherichia coli bacteria). We investigate the changes of surface area to volume ratio of copper nanoparticles prepared in two different methods and its effects on antibacterial activities. We note that slight change of surface area to volume ratio results in the enhancement of its antibacterial activities.*

___________________________________________________________________________

## INTRODUCTION

Metal nanoparticles, due to their special properties and also small dimensions, find important applications in optical, magnetic, thermal, electronic and sensor devices, SERS (surface enhanced Raman scattering), catalysis, etc. Almost all properties of nanoparticles are due to their small sizes. Over the past few decades, inorganic nanoparticles, whose structures exhibit significantly novel and improved physical, chemical, and biological properties, phenomena, and functionality due to their nanoscale size, have elicited much interest. Nanophasic and nanostructure materials are attracting a great deal of attention because of their potential for achieving specific processes and selectivity, especially in biological and pharmaceutical applications.

Applications for copper nanocrystals include as an anti-microbial, anti-biotic and anti-fungal (fungicide) agent when incorporated in coatings, plastics, textiles, in copper diet supplements, in the interconnect for micro, integrated circuits, for its ability to absorb radioactive cesium, in super strong metals, alloys, in nanowire, nanofiber, in certain alloy and catalyst applications.

Further research is being done at various levels for their potential electrical, dielectric, magnetic, optical, imaging, catalytic, biomedical and bioscience properties. Copper nanoparticles are generally immediately available in most volumes. Researchers have also recommended the use of silver and copper ions as superior disinfectants for wastewater generated from hospitals containing infectious microorganisms.

Although only a few studies have reported the antibacterial properties of copper nanoparticles, they show copper nanoparticles have a significant promise as bactericidal agent. However, other nanoparticles, such as platinum, gold, iron oxide, silica and its oxides, and nickel have not shown bactericidal effects in studies with *Escherichia coli*.

**Copper Nanoparticles**

We synthesize copper Nanoparticles for these antibacterial activities studies in accordance with our earlier literature procedure (T.Theivasanthi and M.Alagar, 2010) by dissolving copper sulphate salt in distilled water and electrolyzed. The copper Nanoparticles are formed at the cathode and they are removed carefully. Particle size is noted as 24 nm from XRD analysis by using Debey-Scherrer formula (Instrumental broadening).

$$D = \frac{0.9\lambda}{\beta cos\theta} \quad \ldots\ldots\ldots\ldots\ldots\ldots\ldots (1)$$

Where '$\lambda$' is wave length of X-Ray (0.1541 nm), '$\beta$' is FWHM (full width at half maximum), '$\theta$' is the diffraction angle and 'D' is particle diameter size. Surface Area, Volume and Surface Area to Volume Ratio are calculated and derived from particle size [5].

**Changes in Surface Area to Volume Ratio**

The actual properties of some nanoparticles like silver are fundamentally different at the nanoscale. One prime example of surface area to volume ratio at the nanoscale is gold as a nanoparticle. At the macroscale, gold is an inert element, meaning it does not react with many chemicals, whereas at the nanoscale, gold nanoparticles become extremely reactive and can be used as catalysts to speed up reactions. As the gold nanoparticle size decreases and the surface area to volume ratio increases, the likelihood of ferromagnetism increases. For example, the plasmon resonance of spherical silver nanoparticles results in the particle's exceptional ability to scatter blue light.

Nanoparticles are special and interesting because their chemical and physical properties are different from their macro counterparts. Nanoparticles have unique properties due to their small size. All nanoparticles regardless of their chemical constituents have surface area to volume ratios that are extremely high. This causes nanoparticles physical properties to be dominated by the effect of the surface atoms and capping agents on the nanoparticles surface. The high surface area-to-volume ratio in nanocrystals can lead to unexpected properties. A particle with a high surface area has a greater number of reaction sites than a particle with low surface area, and thus, results in higher chemical reactivity. High surface area to volume ratio is important for applications such as catalysis. The change in properties of materials is due to increased surface area to volume ratio. Reactions take place at the surface of a chemical or material; the greater the surface for the same volume, the greater the reactivity. The link to nanotechnology is that as particles get smaller; their surface area to volume ratio increases dramatically.

This increased reactivity for surface area to volume ratio is widely taken advantage of in nature; one biological example is the body's digestive system. Within the small intestine, there are millions of folds and sub folds that increase the surface area of the inner lining of the digestive tract. These folds allow more nutrients and chemicals to be absorbed at the same time, greatly increasing our body's efficiency and the rate at which we digest food. Another example a cube of sugar, reacting with water as the water dissolves the outside of the sugar. The same size of sugar cube is cut into many little pieces. Each cut makes new outer surfaces for the water to dissolve. The smaller particles of sugar have same volume but have much more surface area.

In order to disclose the effective factors on their antibacterial activity, many studies have already been focused on the powder characteristics, such as specific surface area, particle size and lattice constant by various researchers. Yamamoto et al. have studied the effect of lattice constant on antibacterial activity of ZnO, resulting in the enhancement in antibacterial activity with the slight increase of the lattice constant [2]. However, it is not yet clear what change in antibacterial activity is expected by changing the Surface Area to Volume Ratio of copper nanoparticles. For studying, changes in Surface Area to Volume Ratio of copper nanoparticles and its effects on antibacterial activities of copper nanoparticles, we have compared Surface Area to Volume Ratio of copper nanoparticles synthesized in electrolysis method with chemical reduction synthesis of copper nanoparticles by N.Prakash et al. [4] and the details are in Table.1.

**Table 1: Comparision of Surface Area to Volume Ratio and antibacterial activities of Copper Nanoparticles on *Escherichia Coli* (Gram Negative bacteria)**

| Copper Nanoparticles synthesis method | Surface Area ($nm^2$) | Volume ($nm^3$) | Surface Area to Volume Ratio | Inhibition Zone diameter (in mm) |
|---|---|---|---|---|
| Electrolysis method | 1809 | 7235 | 0.25 | 15 |
| Chemical Reduction Method | 45216 | 904320 | 0.05 | 8-10 |

From this comparison, antibacterial activities of copper nanoparticles prepared in electrolysis method is more on *Escherichia Coli* (Gram Negative bacteria) than the copper nanoparticles prepared in chemical reduction method. We also noted that the slight change of Surface Area to Volume Ratio results in enhancement of antibacterial activities of copper nanoparticles.

**Copper Nanoparticles and Its Antibacterial Activities**

Nanomaterials are the leading in the field of nanomedicine, bionanotechnology and in that respect nanotoxicology research is gaining great importance. The US Environmental Protection Agency (EPA) has approved registration of copper as an antimicrobial agent which is able to reduce specific harmful bacteria linked to potentially deadly microbial infections (European Copper Institute, 2008). In addition, no research has discovered any bacteria able to develop immunity to copper as they often do with antibiotics. The emergence of nanoscience and nanotechnology in the last decade presents opportunities for exploring the bactericidal effect of metal nanoparticles. The bactericidal effect of metal nanoparticles has been attributed to their small size and high surface to volume ratio, which allows them to interact closely with microbial membranes and is not merely due to the release of metal ions in solution.

A cell wall is present around the outside of the bacterial cell membrane and it is essential to the survival of bacteria. It is made from polysaccharides and peptides named *peptidoglycon*. There are broadly speaking two different types of cell wall in bacteria, called gram-positive and gram-negative. The names originate from the reaction of cells to the gram stain, a test long-employed for the classification of bacterial species. Gram-positive bacteria possess a thick cell wall containing many layers of peptidoglycan. In contrast, gram-negative bacteria have a relatively thin cell wall consisting of a few layers of peptidoglycan. Surfaces of copper nanoparticles affect / interact directly with the bacterial outer membrane, causing the membrane to rupture and killing bacteria.

Antibacterial activity of copper nanoparticles synthesized by electrolysis was evaluated by using standard Zone of Inhibition (ZOI) microbiology assay. The sample copper nanoparticles prepared in electrolysis method showed diameter of inhibition zone against *E.Coli* 15 mm and *B.megaterium* 5 mm. The results are shown in Figures 1 and 2.

**Antibacterial activities evaluation of Electrolytic synthesized Copper nanoparticles (sample no.1)**

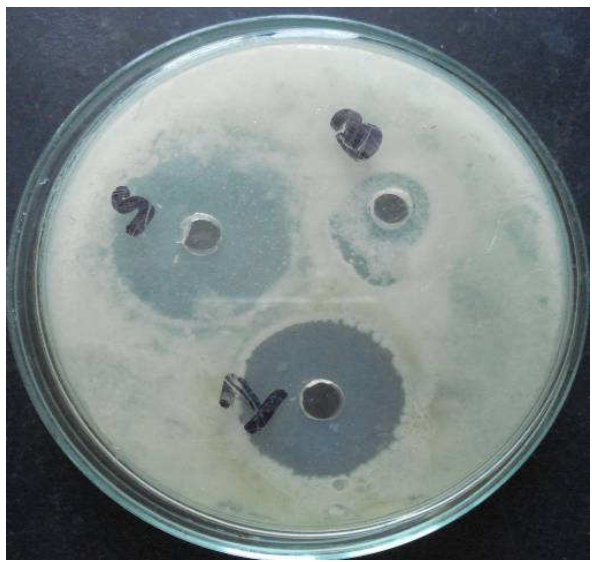

**Fig. 1. Zone of inhibition diameter against *Escherichia coli* bacteria - 15 mm**

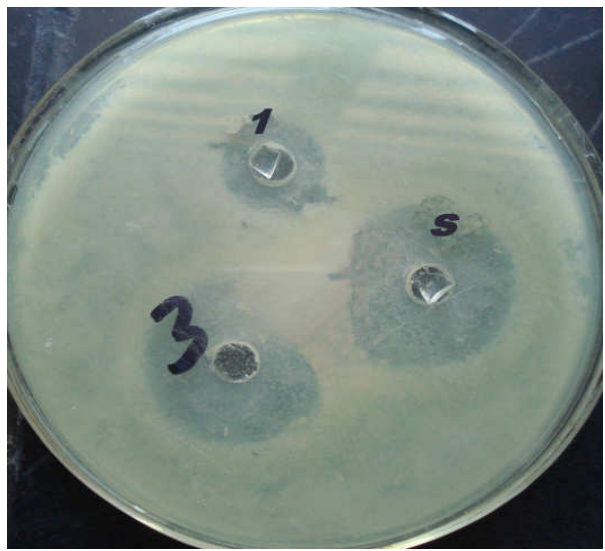

**Fig. 2. Zone of inhibition diameter against *Bacillus megaterium* - 5 mm**

**Enhancement of Antibacterial Activities**

People have used copper for its antibacterial qualities for many centuries. However, copper nanoparticles have showed antibacterial activities more than copper. Various researchers have tried to enhance the antibacterial actions of metals and metal oxides nanoparticles adopting various methods i.e. using capping agents while on synthesis, using a combination of light energy with nanoparticles, using a combination of ultrasound wave with nanoparticles, using a combination of electric field with nanoparticles etc.

D.K.Tiwari and J. Behari reported that the silver nanoparticles treated with short time exposure with ultrasound show increased antibacterial effect but this time was not enough to kill the bacteria with ultrasound only [1]. It indicated that synergistic effect of ultrasound & silver nanoparticles. The ultrasound facilitates the entry of Ag-nanoparticles inside the cells and the antibacterial effect was enhanced with same concentration of nanoparticles in presence of ultrasound waves. The biocidal effect was more pronounced when compared to the actions of silver nanoparticles alone.

Omid Akhavan and Elham Ghaderi investigated the effect of an electric field on the antibacterial activity silver nanorods against *E. coli* bacteria [3]. It was found that the grown silver nanorods show strong and fast antibacterial activity. Applying an electric field in the direction of the nanorods (without any electrical connection between the nanorods and the capacitor plates producing the electric field) promoted their antibacterial activity. This indicated that the antibacterial activity of silver nanorods can be enhanced by applying an electric field.

In view of the above, we have tried to increase antibacterial activities of copper nanoparticles for which we have made an attempt using electrical power while on synthesizing of copper nanoparticles. We have synthesized copper nanoparticles in electrolysis by using electrical power. Copper nanoparticles synthesized in this method have showed more antibacterial activities (For *E.Coli* bacteria) than copper nanoparticles synthesized in chemical reduction method and the details are in Table.2.

**Table.2. Comparision of activities of Copper Nanoparticles on Gram (-) and Gram (+) bacteria**

| Copper Nanoparticles synthesis method | Name of Bacteria | Variety of Bacteria | Inhibition Zone diameter (in mm) |
|---|---|---|---|
| Electrolysis method | *Escherichia coli* | Gram (-) | 15 |
| | *Bacillus megaterium* | Gram (+) | 5 |
| Chemical Reduction method | *Staphylococcus aureus* | Gram (+) | 6-8 |
| | *Salmonella typhimurium* | Gram (-) | 6-8 |
| | *Pseudomonas aeruginosa* | Gram (-) | 6-8 |
| | *Escherichia coli* | Gram (-) | 8-10 |

## RESULTS AND DISCUSSIONS

Surface Area to Volume Ratio of copper nanoparticles synthesized in electrolysis method and chemical reduction method was calculated and compared. We noted that changes in Surface Area to Volume Ratio of copper nanoparticles were showing more antibacterial activities. We also noted that using electrical power while on synthesizing of copper nanoparticles is increasing its antibacterial activities. Actions of copper nanoparticles synthesized in both above method was also observed against both gram (-) and gram (+) bacteria.

## CONCLUSION

We have come for conclusion that copper nanoparticles synthesized in electrolysis method are showing antibacterial activities against both gram (-) and gram (+) bacteria. Changes in Surface Area to Volume Ratio of copper are enhancing its antibacterial activities. Copper nanoparticles synthesized in electrolysis method are showing more antibacterial activities (For *E.Coli* bacteria) than copper nanoparticles synthesized in chemical reduction method. Using electrical power while on synthesizing of copper nanoparticles is increasing its antibacterial activities. The chemicals involved in the synthesis of nanoparticles are commonly available, cheap, and non-toxic. The technology can be implemented with minimum infrastructure. The experiments suggest the possibility to use this material in water purification, air filtration, air quality management, antibacterial packaging, etc.

**Acknowledgements**

The authors express immense thanks to **Dr.G.Venkadamanickam**, *Rajiv Gandhi Cancer Institute & Research Center,* Delhi, India, **Dr.M.Palanivelu**, *Arulmigu Kalasalingam College of Pharmacy (Kalasalingam University,* Krishnankoil, India), Staff & Management of *PACR Polytechnic College*, Rajapalayam, India and *Ayya Nadar Janaki Ammal College*, Sivakasi, India for their valuable suggestions, assistances and encouragements.